\documentclass[11pt,preprint]{aastex}

\slugcomment{To appear in 2006/12/01 issue of ApJ}
\shorttitle{Chromatic Shear Effect on ExAO}
\shortauthors{Nakajima}

\begin{document}

\title{Zenith-Distance Dependence of Chromatic Shear Effect: \\
A Limiting Factor for an Extreme Adaptive Optics System}

\author{Tadashi Nakajima}
\affil{National Astronomical Observatory of Japan \\ 
Osawa 2-21-1, Mitaka, 181-8588, 
Japan
}

\begin{abstract}
Consider a perfect AO system with a very fine
wavefront sampling interval and a very small actuator interval.
If this AO system
senses wavefront at a wavelength, $\lambda_{WFS}$,
and does science imaging at another wavelength, $\lambda_{SCI}$,
the light paths through the turbulent atmosphere
at these two wavelengths are slightly different
for a finite zenith distance, $z$. The error in wavefront  
reconstruction of the science channel associated with this non-common path effect,
or so-called chromatic shear, 
is uncorrectable and sets an upper bound of the system performance.
We evaluate the wavefront variance, $\sigma^2(\lambda_{WFS},\lambda_{SCI},z)$
for a typical seeing condition at Mauna Kea and
find that this effect is not negligible at a large $z$.
If we require that the Strehl ratio be
greater than 99 or 95\%, $z$ must be less than about 
50 or 60$^\circ$ respectively,
for the combination of visible wavefront sensing and infrared science 
imaging. 
\end{abstract}


\section{Introduction}

It is well known that a high-performance adaptive optics
system with a Strehl ratio well over 90\% is required 
for the direct detection of an exoplanet from the ground especially
in reflected light. Such an adaptive optics system is often called 
an extreme AO system (ExAO), because of its high level of sophistication
in instrumentation. The ExAO has strict requirements on guide stars and
wavefront sensors.
Fine wavefront sampling is achieved by many pixels in the
wavefront sensor, each of which
corresponds to the length scale less than cm.
Fine temporal sampling is also required along with the fine spatial sampling.
Therefore, a large number of photons are required for a tiny area 
on the wavefront for a short period of time. 
So an AO guide star must be very bright. Another requirement on
the guide star is that it must be regarded as a point source, even after the AO
correction,
and a presently available laser guide star, located at a finite altitude
with a finite angular extent, is not fitted for this purpose,
even if the problem of brightness or laser power is overcome.
If the Strehl reduction due to the finite extent of the laser guide star
is inevitable, there is no point of sampling the wavefront so finely,
in other words, there is no use of an ExAO.
For the same reason, a stringent requirement on isoplanicity will
result in a small field of view.
For these reasons, the major application of an ExAO requires
a very bright natural guide star and the high Strehl ratio
is achieved only for a small field of view.
All things considered, an exoplanet search around bright nearby stars,
will remain the primary scientific goal for an ExAO project.

There are many sources of errors that can cause the reduction of the Strehl
ratio, e.g. see \citep{guyon}, but most of them are in principle controllable
by sophisticated instrumentation. 
However, there are also some 
errors which are in principle uncorrectable. 
Here we focus on one of the major uncorrectable errors: atmospheric chromatic non-common-path
error, or ``chromatic shear (CS)", associated with the different paths through the
turbulent atmosphere of the light beams incident on the visible wavefront sensor and
the infrared science imager, when the target star is at a finite zenith distance.

In actual astronomical observations, total observing time for
an object is finite and the object never stays at the zenith.
Moreover, it often happens that an object needs to be observed at a large zenith
distance, due a large difference between the declination of
the object and the latitude of the observatory.
In order to secure a significant amount of observing time
for an exoplanet search, it is inevitable to observe a target star
with a finite zenith distance.

Most of the current AO systems sense wavefront at visible
wavelengths and image the target at near-infrared wavelengths.
This is partly due to the higher availability of a visible detector
such as a CCD
for a wavefront sensing camera.
However, if the CS effect is a significant
limiting factor for the next-generation extreme AO systems whose
scientific justification is direct exoplanet detection,
the choice of the combination of $\lambda_{WFS}$ and $\lambda_{SCI}$,
may need to be reconsidered.

In this paper, we quantify the magnitude of 
this CS effect, in terms of
wavefront variance associated with it.

\section{Formulation}

\subsection{Refractive Index and Atmospheric Model}

According to \citet{Cox}, using
the index of refraction, $n_s(\lambda)$, for dry air 
at a standard pressure, $p_s$, a standard absolute temperature, $T_s$,
the refractive index $n$ for the pressure, $p$, and absolute temperature,
$T$, is given by,

\begin{equation}
n(\lambda; p,t) - 1 = \left(\frac{p}{p_s}\right)\left(\frac{T_s}{T}\right)
(n_s(\lambda)-1).
\end{equation}

Strictly speaking, $n$ is also dependent on the water vapor pressure. However
its effect is a minor reduction of $n$ and the dry air case represents the 
worst case designing.
To obtain the light path through the atmosphere,
we need to know the vertical profiles of $p(h)$ and $T(h)$,
where $h$ is altitude. We adopt an atmosphere model
of the Earth used in the field of aeronautics, in which 
the pressure and temperature are given as analytic functions of
the altitude, $h$ (http://www.grc.nasa.gov/WWW/K-12/airplane/atmosmet.html).
This model covers from the troposphere ($h > 0$ km) to
the stratosphere ($h < 50$ km), while 
turbulent layers of significance for optical wave propagation is
located at altitudes below the lower stratosphere ($h < 25$ km). 
Then the refractive index $n$ becomes a function of
$\lambda$, and $h$.

If the zenith distance of a star incident on the upper atmosphere from
space is
$z_0$, the apparent zenith distance, $z(\lambda,h)$, is related to $z_0$
by

\begin{equation}
n(\lambda,h) \cdot \sin z(\lambda,h) = \sin z_0.
\label{snell3}
\end{equation}

At an arbitrary altitude in the atmosphere, $h$, Snell's law can be expressed in a differential form as,

\begin{equation}
dn(\lambda,h) \cdot \sin z(\lambda,h) + n(\lambda,h) \cdot \cos z(\lambda,h) \cdot dz = 0.
\label{snell-1}
\end{equation}

or 

\begin{equation}
dz(\lambda,h)  = -\tan z(\lambda,h) \cdot dn(\lambda,h) / n(\lambda,h).
\label{snell-2}
\end{equation}

In principle, we must integrate this differential relation (\ref{snell-2}) along the light path.
However, for the practical purpose,
a more useful approximate expression for
the bending angle, $\Delta \theta(\lambda,h)$,
not in a differential form, is obtained, since 
the index of refraction of the air is very close to that of the vacuum
and the total bending angle is also small.
By taking the first order terms of eq(\ref{snell3}),

\begin{equation}
\Delta \theta(\lambda,h) = -\tan z_0 \cdot (n(\lambda,h)-1).
\label{snell4}
\end{equation}

To evaluate CS, we need to
know the separation of the two beams at 
$\lambda_{WFS}$ and $\lambda_{SCI}$, perpendicular
to the line of sight in length, $r$,
at a given altitude, $h$ and zenith distance, $z_0$.

In actual observing with an AO system, an atmospheric dispersion
corrector will be used and the two beams are
aligned inside the optical system. The two beams gradually
separate towards the upper atmosphere and eventually become 
parallel above the stratosphere.

Since the angle between the two beams, $\alpha$ is given by

\begin{eqnarray}
\alpha(\lambda_{WFS},\lambda_{SCI},h) & = & \Delta \theta(\lambda_{WFS},h) 
- \Delta \theta(\lambda_{SCI},h)  \nonumber \\
& = & -\tan z_0 \cdot (n(\lambda_{WFS},h)-n(\lambda_{SCI},h)). 
\label{alpha}
\end{eqnarray}

We calculate $r$ as follows.
Let $H_T$ and $H_{MAX}$ be the altitude of the telescope
and the upper limit of the atmosphere.
Then a relative ray displacement due to differential refraction,
$r$ is obtained by the following integration,

\begin{equation}
r(\lambda_{WFS},\lambda_{SCI},h) = \int_{H_T}^{h}
\alpha(\lambda_{WFS},\lambda_{SCI},h) / \cos z_0 dh. 
\label{integral}
\end{equation}

\subsection{Turbulence Model and Wavefront Variance}

The turbulence profile 
is characterized
by the refractive index structure constant, $C_n^2(h)$,
whose integral is related to the Fried parameter, $r_0$ \citep{fried66}, 
by

\begin{equation}
r_0 = 
0.185 
\left[\frac{\lambda_{SCI}^2}{\int_{H_T}^{H_{MAX}} C_n^2(h) dh}\right]^{3/5}.
\label{fried}
\end{equation}
 
In our calculation, a seeing condition is specified by $r_0$ at a given
wavelength,  $\lambda_{SCI}$. Then from eq(\ref{fried}), the integral of
$C_n^2$, $\int_{H_T}^{H_{MAX}} C_n^2(h) dh$ is obtained.
For a good seeing condition typical of an astronomical site,
the variation of 
$C_n^2(h)$ as a function of $h$ is slow. Under such conditions,
the wave propagation through a turbulent medium can be
treated by ``smooth perturbations'' \citep{tatarskii}, 
which give the overall wavefront structure function, $D(r)$,
as the sum of the wavefront structure functions of individual
turbulent layers as

\begin{equation}
D(r) = \sum_i D_i(r),
\label{sf}
\end{equation}

where $D_i(r)$ is the structure function of the $i^{\rm th}$ layer.
It is also assumed that the wavefront error is mostly in phase,
not in amplitude or scintillation, which should be valid 
at a representative astronomical site like Mauna Kea.
Therefore $D(r)$ can be regarded as a phase structure function and

\begin{equation}
D_i(r) = 2.91 \left(\frac{2 \pi}{\lambda_{SCI}}\right)^2 \cdot {\int_{i} C_n^2(h) dh} \cdot r^{5/3},
\end{equation}

where $\int_i dh$ indicates the integral over the $i^{\rm th}$ turbulent layer.
We adopt the six-layer turbulence model employed by \citet{guyon}
to approximate the conditions above Mauna Kea (Table 4 of Guyon 2005).
The $i^{\rm th}$ turbulent layer is characterized by the altitude, $h_i$,
and the fractional contribution to the integrated  $C_n^2(h)$, $F_i$, or,

\begin{equation}
F_i = \frac{\int_{i} C_n^2(h) dh}
{\int_{H_T}^{H_{MAX}} C_n^2(h) dh},
\end{equation}

where the physical thickness of the 
$i^{\rm th}$ layer is negligible compared to
the entire thickness of the atmosphere. 

The wavefront variance
associated with this layer is,

\begin{eqnarray}
\sigma_i^2 & = & D_i(r(\lambda_{WFS},\lambda_{SCI},h_i)) \nonumber \\
& = &
2.91 \left(\frac{2 \pi}{\lambda_{SCI}}\right)^2 
F_i \cdot {\int_{H_T}^{H_{MAX}} C_n^2(h) dh} \cdot
r(\lambda_{WFS},\lambda_{SCI},h_i)^{5/3} / \cos z_0,
\end{eqnarray}

Note that the factor $1/ \cos z_0$ is necessary, since the
integrated $C_n^2$ is given for the propagation toward the
zenith, while the actual light path through
the layer of isotropic turbulence is longer by this factor. 
Here the small variance caused by the difference in the optical path lengths at the two
wavelengths is neglected.
We obtain the total variance $\sigma^2(\lambda_{WFS},\lambda_{SCI},z_0)$
by

\begin{equation}
\sigma^2(\lambda_{WFS},\lambda_{SCI},z_0) = \sum_{i=1}^{6} \sigma_i^2.
\end{equation}

Computer programming is carried out in C, and source codes are
available upon request to the author. As long as the multi-layer turbulence
model is adopted, they can be modified easily for other astronomical sites.

\section{Results}

Typical situations are that $\lambda_{WFS}$ is in the visible 
and $\lambda_{SCI}$ is in the near-infrared.
In addition, we also explore the possibility of using an infrared wavefront sensor.

For a typical CCD detector, the quantum efficiency (QE) peaks around 0.7
$\mu$m, while a full-depletion type CCD with high QE at longer wavelengths
may be used at around 0.9 $\mu$m \citep{groom00,strueder00,kamata04}. 
The infrared wavefront sensor must be
operated through atmospheric windows, namely at $J$ (1.25 $\mu$m),
$H$ (1.65 $\mu$m), or $K$ (2.2 $\mu$m). One potential disadvantage 
in the infrared is that sky background is much higher 
than in the visible, especially at $H$ due to OH airglow and
at $K$ due to both OH airglow and thermal emission.
Although we include the case for wavefront sensing at $H$ and
science imaging at $K$, a realistic long wavelength limit may be 
1.25 $\mu$m, where sky background is still tolerable.
To summarize, we consider the cases for 
$\lambda_{WFS} =$ 0.7, 0.9, 1.25, and 1.65 $\mu$m,
and $\lambda_{SCI}$ = 1.25, 1.65, and 2.2 $\mu$m.

For the seeing condition, we choose a typical value for
Mauna Kea, $r_0 = 0.2$ m at $0.5 \mu$m, which corresponds to
a half arcsecond seeing in the visible. Since the wavefront
variance scales as $(0.2/r_0)^{5/3}$, it is sufficient to
present the results only for a single value of $r_0$.
However, when we need to estimate the variance for a site other than Mauna Kea,
this scaling does necessarily hold, since the altitudes and
strengths of individual turbulent 
layers determine the overall wavefront variance.

Results are plotted as the zenith angle vs. $\log_{10} \sigma^2$,
for $\lambda_{SCI} = 1.25, 1.65$, and 2.2 $\mu$m 
respectively in Figs 1, 2, and 3. 
For another seeing condition characterized by $r_0$,
variances are obtained by shifting individual curves by $(5/3) \log_{10} (0.2/r_0)$. 
Since the Strehl ratio, $S$, is given by

\begin{equation}
S = \exp ( -\sigma^2 ),
\end{equation}

or for a small $\sigma^2$,

\begin{equation}
S \approx 1 -\sigma^2,
\end{equation}

\citep{Tyson1991}.

Therefore,
$\log_{10} \sigma^2 = -2$ and -1, correspond to $S = 0.99$ and 0.90 respectively.

Another way of presenting our results is to show
the behavior of  the maximum allowed zenith distance 
as a function of $\lambda_{WFS}$, for a given set of
$\lambda_{SCI}$  and the tolerance level of the wavefront
variance.
In Figs 4, 5 and 6, such plots are given for $\lambda_{SCI}$ = 1.25, 1.65, and 2.2 $\mu$m,
for the cases of $\sigma^2$ = 0.01 and 0.05.

\section{Discussion}

\subsection{Significance of High Altitude Turbulence}

To see the relative contribution of each layer, its altitude
relative to the summit of Mauna Kea (4.2 km), the beam separation $r_i$
in cm, the phase variance, $\sigma_i^2$ in rad$^2$ are given
for $\lambda_{SCI}=1.65$ $\mu$m, $\lambda_{WFS} = 0.7$ $\mu$m, 
and $z=50^\circ$ in Table 1.
The total wavefront variance for this case is about 0.01.
Two factors that determine the relative significance of each layer, are
the fractional integrated $C_n^2$, $F_i$, and beam separation $r_i$, since
$\sigma_i^2 \propto F_i \cdot r_i^{5/3}$. 
The largest contribution comes from the layer 8 km above the Mauna Kea Observatory.
$r_i$ at a high altitude is order of a few cm. 

\subsection{Is the Chromatic Shear a Limiting Factor 
for the Next Generation Extreme AO System?}

If we consider a Shack-Hartman sensor with a 512$\times$512 pixel CCD,
attached to an 8 m telescope, the wavefront sampling interval is 1.6 cm, comparable
to $r_i$. On the other hand, an ExAO currently designed will have
about 2000 actuators, or 50 actuators
across the diameter of the telescope aperture, corresponding to 16 cm sampling interval.
The residual wavefront variance, $\sigma^2_{AO}$  associated
with the finite actuator spacing  will be given by

\begin{equation}
\sigma^2_{AO} = 0.1 C \left(\frac{0.16}{r_0}\right)^{5/3},
\end{equation}

where the constant $C$ depends on the actual 2D configuration of
the actuator, and
is order of unity, assuming that the tip/tilt component of the
wavefront error is removed on
the spatial scale of the actuator interval. Even if some other low-order
Zernike terms are corrected, there will not be a major change in $C$,
since the major part of the wavefront variance is due to the tip/tilt effect
\citep{noll}. 
For instance, $C = 1.34$ for a 16-cm-diameter tip/tilt-corrected circular aperture \citep{noll}. 
For a finite zenith distance,  

\begin{equation}
\sigma^2_{AO}(z) = 0.1 C \left(\frac{0.16}{r_0(\lambda_{SCI})}\right)^{5/3} / \cos z,
\end{equation}

since the path length through the turbulent atmosphere is longer by $1/\cos z$.
When the seeing condition is characterized by  $r_0 = 0.2$ $\mu$m, 
at $\lambda = 0.5$ $\mu$m,

\begin{equation}
r_0(\lambda_{SCI}) = 0.2 \cdot \left(\frac{\lambda_{SCI}}{0.5}\right)^{1.2}.
\end{equation}

\begin{equation}
\sigma^2_{AO}(z) = 0.0172 C \frac{1}{\lambda_{SCI}^2 \cdot \cos z},
\end{equation}

At $z = 50^\circ$, $\sigma^2_{AO} = 0.017C, 0.0098C$, and $0.0055C$ respectively
at $\lambda_{SCI} = 1.25, 1.65$ and $2.2$ $\mu$m.
Remember that $\sigma^2 = 0.01$ was obtained for  
$\lambda_{SCI} = 1.65$ $\mu$m,
$\lambda_{WFS} = 0.7$ $\mu$m, $z = 50^\circ$ for the pure CS effect
given in Table 1.
Therefore at this zenith distance, wavefront errors originating from 
the finite actuator spacing (FAS)  
and the CS effect appear comparable.
However $z$-dependencies of two effects are different.
The variance due to the CS effect approximately depends on
$z$ as $(\sin z)^{5/3} \cdot (1/\cos z)^{13/3}$,
while that due to the FAS effect goes as $1/\cos z$.
Therefore the former is a much faster increasing function
of $z$.
At $z<50^\circ$, the FAS effect is dominant, while at $z > 50^\circ$,
the CS effect becomes dominant.
This parameter combination is of practical interest, since
$\lambda_{WFS} = 0.7$ $\mu$m 
represents the operational wavelength of a commonly used
wavefront sensor and  $\lambda_{SCI} = 1.65$ $\mu$m is considered to be
the prime wavelength band for simultaneous differential imaging aiming
at detecting giant planets having methane bands 
using a room-temperature coronagraph.

\subsection{Spatial Frequency Dependence of the Chromatic Shear}

In the previous subsection, we compared only the total wavefront variances due
to the CS and FAS effects. However, the spatial power spectra of the two effects
are expected to be different and
here we discuss this difference qualitatively.

The CS effect preserves the original Kolmogorov spectrum, including low frequency
components. Therefore at the focal plane, it degrades the target star PSF
at small angular radii. So the detectability of a close-in planet is 
reduced by the CS effect. 

On the other hand, the AO correction works as a high-pass filter of the Kolmogorov spectrum,
and the residual power spectrum due to the FAS effect has finite values mainly at high frequencies.
At the focal plane, the PSF envelope at a radius, 2$^{\prime\prime}$ or greater
is affected for the 8 m telescope at  $\lambda_{SCI} = 1.65$ $\mu$m.
High frequency phase fluctuations produce extended scattered light at the focal plane,
which is in competition with OH sky background of 13.4 mag per $\square^{\prime\prime}$ \citep{Cox}.

\subsection{More on the Next Generation
Extreme AO System}

From the previous subsection, we have seen that there will be a 
transition zenith distance, $z_{TR}$, at which the wavefront variance due to the FAS effect
dominating the wavefront error at $z < z_{TR}$ and that due to the CS effect
dominating the wavefront error at  $z > z_{TR}$, cross over.
Here we wish to shift $z_{TR}$ as large as possible.
The quantitative optimization process depends on the actual design of an ExAO system, whose consideration
is beyond the scope of this paper. Here we only list up some issues to be considered.

A simple-minded solution is to shift $\lambda_{WFS}$ to a longer wavelength.
However many practical issues arise in reality. 

The first issue is
sky background, which we have already mentioned before. Actually
a smaller  $\lambda_{WFS}$ is preferred, if this effect is dominant.
Only planets with large angular separations from their main stars are
affected by sky background. Depending on the Strehl ratio of
the system, the FAS effect, which is larger for a smaller $\lambda_{WFS}$,  may dominate the background instead of
the sky background as mentioned above.

The second issue is related to the detector performance, whose situation
may change in the near future. It is not the quantum efficiency, which
is already high enough both in the visible and  near infrared,
but the read-noise level, which can be a significant issue,
due to the fast temporal sampling of the wavefront and a small collecting area
per actuator. A currently available infrared array typically has a
read-noise level of about 10 $e^{-}$ per pixel, and may even be higher
for a higher sampling rate.  Actually at this read noise level,
it is not sky background, but the read noise that determines the signal-to-noise ratio.
Unless a low-noise infrared detector happens to become available,
it may be realistic to choose a visible detector.

One may try to optimize the performance of the wavefront sensor by
the combination of a red-sensitive CCD and a long-pass filter to
reject blue photons, which behave as a noise source when 
the CS effect is dominant, provided that the red-sensitive CCD shows
a low read-noise level even for a high sampling rate.
However the quantitative analysis should take into
account many factors including the spectral energy distribution of a target star,
and further general discussion may not be of a major practical use.

We might have underestimated the instrumental error associated with
a non-ideal adaptive optics system. So the actual instrumental error
may be larger than that estimated from the FAS effect alone.
If this is the case, a true $z_{TR}$ will be larger than that estimated above.

\section{Concluding Remark}
 
We have obtained the wavefront variance due to atmospheric non-common path error,
or chromatic shear, 
as a function of wavelengths for wavefront sensing and science imaging,
and zenith distance,
assuming an idealized adaptive optics system with a very fine 
actuator interval for wavefront correction.
For visible wavefront sensing and infrared science imaging,
the wavefront variance exceeds 0.01 at $z \sim 50^\circ$.
For an extreme AO system currently designed, the residual error
due to the finite actuator spacing, appears to dominate
the wavefront error at $z < 50^\circ$ and the non-common path effect
may become dominant only at $z > 50^\circ$. 







\acknowledgments

The author thanks Olivier Guyon and the referee, Eugene Pluzhnik, for helpful comments on the manuscript
of this paper.

\clearpage

\begin{table}
\begin{center}
\begin{tabular}{ccccc}
\tableline \tableline
Layer & Altitude & Fractional $\int C_n^2 dh$  & Beam Separation & Wavefront Variance \\
      &    km (above Mauna Kea)    &                   &      cm         &       rad$^2$  \\
\tableline
1    &  0.5  & 0.2283 & 0.17 & 7.98$\times 10^{-5}$ \\
2    & 1.0   & 0.0883 & 0.30 & 8.14$\times 10^{-5}$ \\
3    & 2.0   & 0.0666 & 0.57 & 1.78$\times 10^{-4}$ \\
4    & 4.0   & 0.1458 & 1.04 & 1.04$\times 10^{-3}$ \\
5    & 8.0   & 0.3350 & 1.68 & 5.34$\times 10^{-3}$ \\
6    & 16.0  & 0.1350 & 2.22 & 3.42$\times 10^{-3}$ \\
\tableline
Total &      &   1.0   &     & 1.01$\times 10^{-2}$ \\
\tableline      
\end{tabular}
\caption{Relative Contribution of Each Turbulent Layer. $\lambda_{SCI}=1.65$ $\mu$m,
$\lambda_{WFS}=0.7$ $\mu$m and $z=50^\circ$. This six-layer model is 
adopted from \citet{guyon}.}
\end{center}
\end{table}




\clearpage

\begin{figure}[tbp]
\includegraphics[angle=-90]{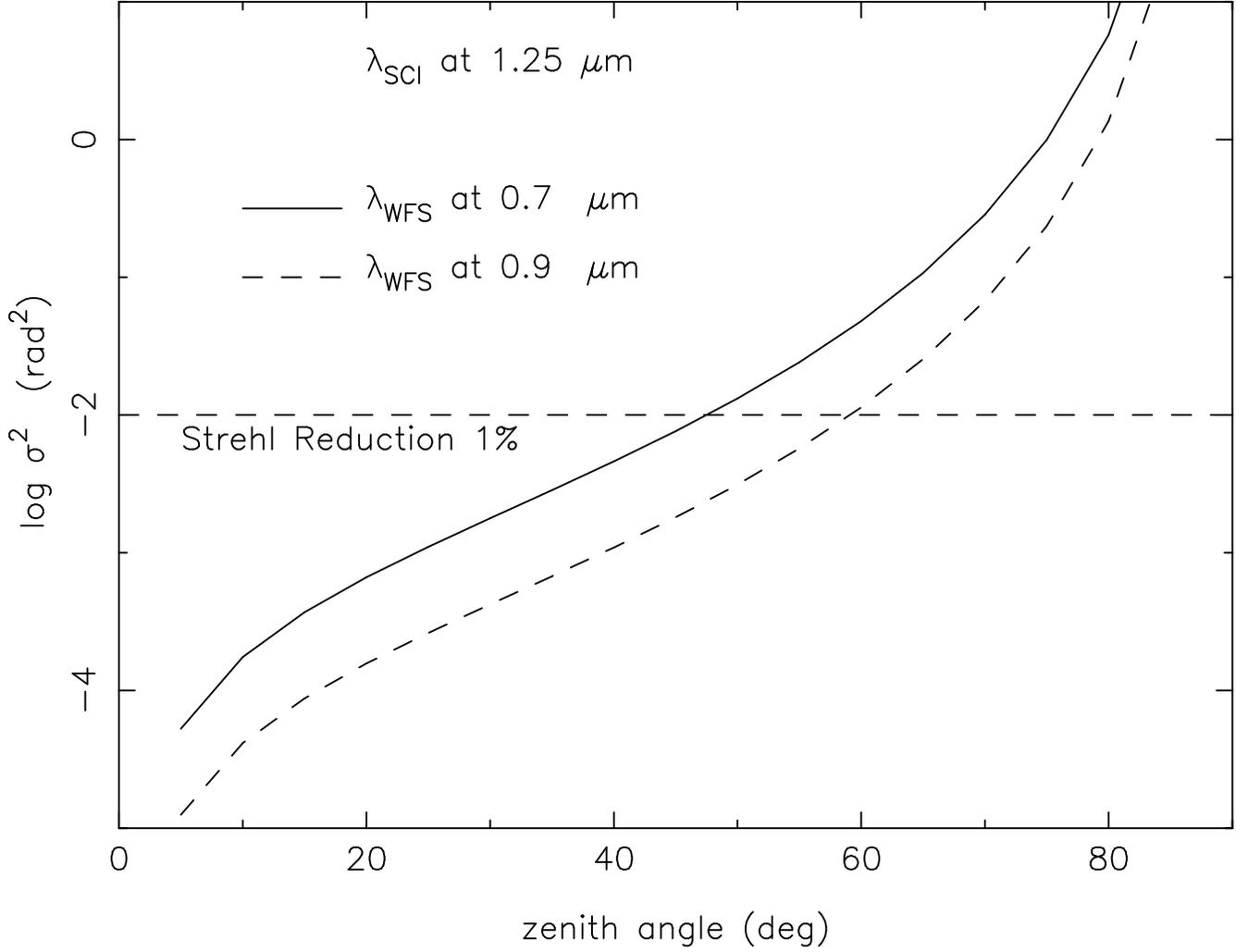}
\caption{Wavefront variance due to atmospheric non-common path effect.
$\lambda_{SCI} = 1.25 $ $\mu$m.
Assumed seeing condition is $r_0 = 0.2$m at $\lambda=0.5 $ $\mu$m.
The variance for an arbitrary $r_0$ can be obtained by 
shifting each curve vertically by $(5/3)\log_{10}(0.2/r_0)$.
\label{S1.25}}
\end{figure}

\clearpage

\begin{figure}[tbp]
\includegraphics[angle=-90]{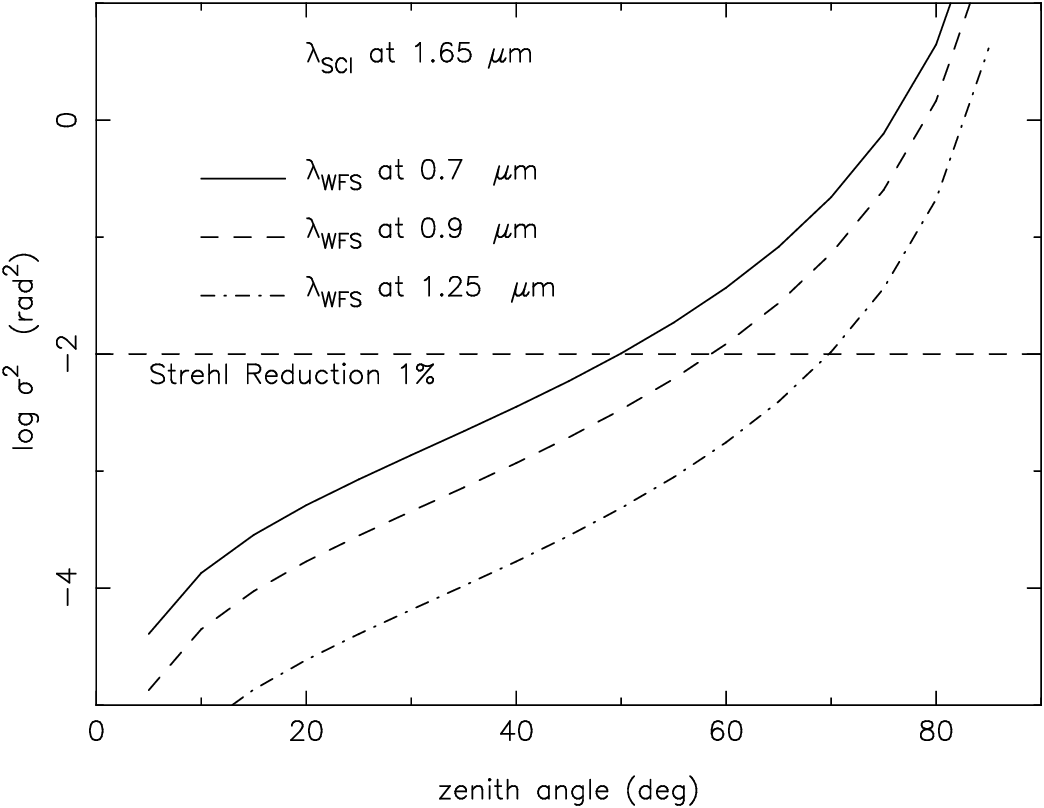}
\caption{Wavefront variance due to atmospheric non-common path effect.
$\lambda_{SCI} = 1.65 $ $\mu$m.
\label{S1.65}}
\end{figure}

\clearpage

\begin{figure}[tbp]
\includegraphics[angle=-90]{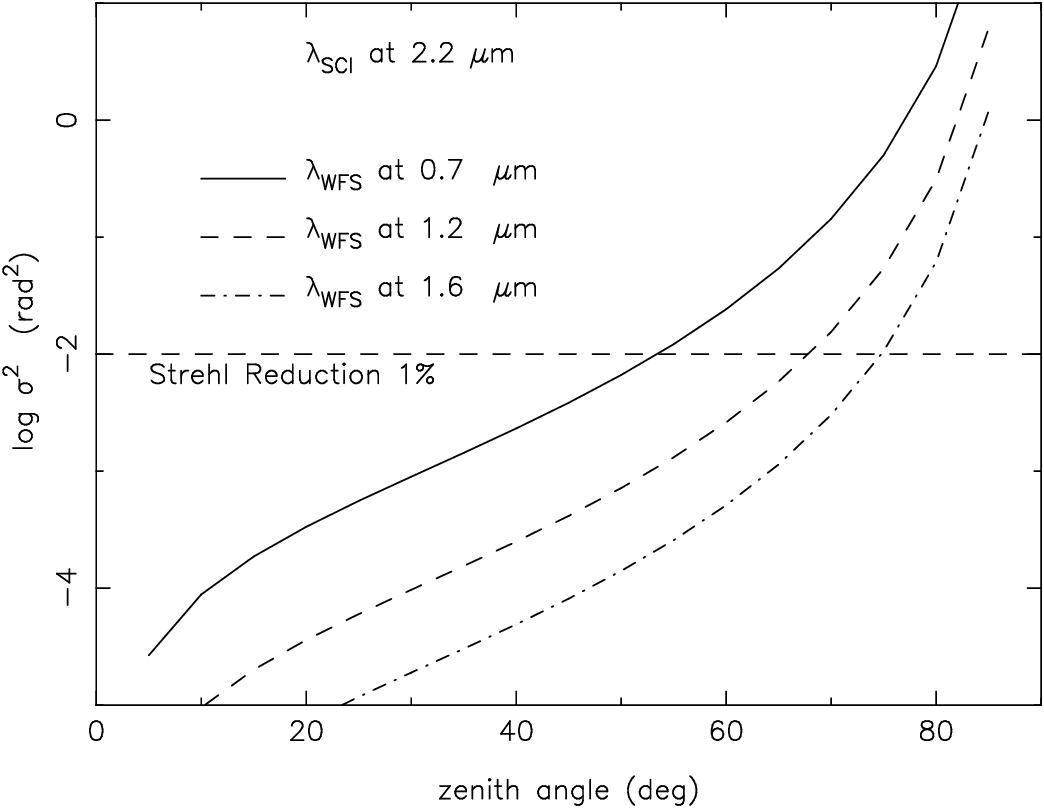}
\caption{Wavefront variance due to atmospheric non-common path effect.
$\lambda_{SCI} = 2.2 $ $\mu$m.
\label{S2.2}}
\end{figure}

\clearpage

\begin{figure}[tbp]
\includegraphics[angle=-90]{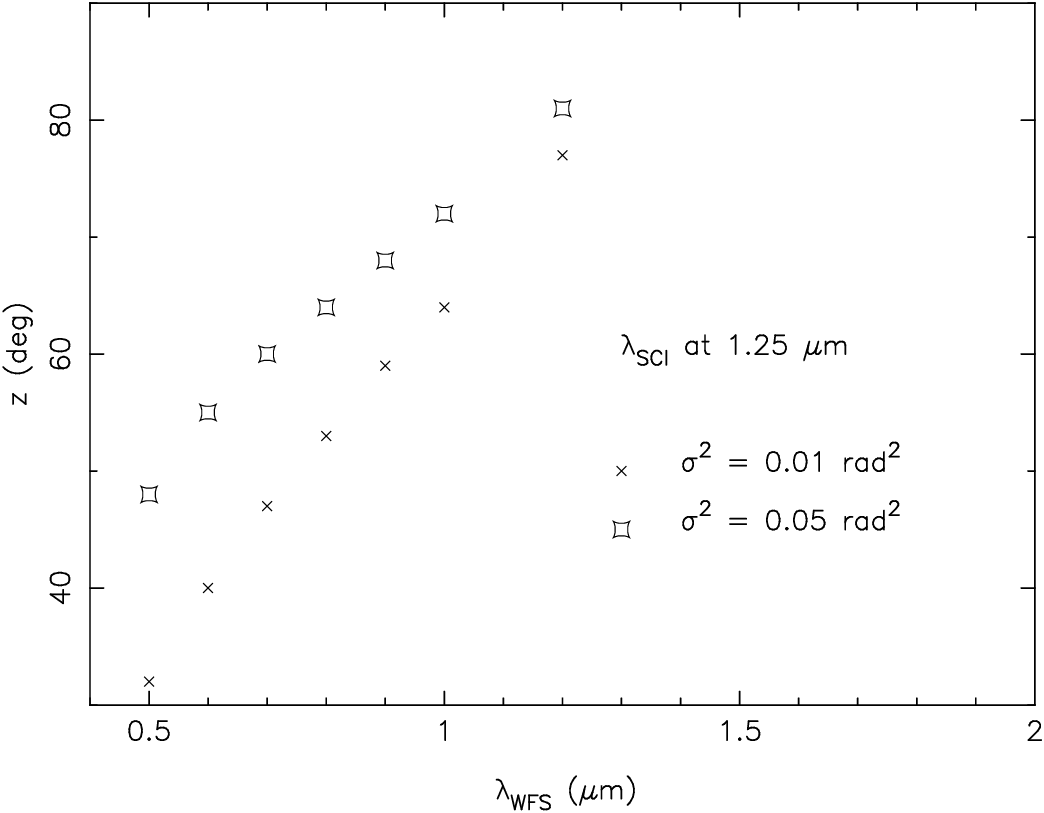}
\caption{Wavelength dependencies of zenith distances giving 
$\sigma^2$ = 0.01 and 0.05.  $\lambda_{SCI} = 1.25 $ $\mu$m.
\label{Z1.25}}
\end{figure}

\clearpage

\begin{figure}[tbp]
\includegraphics[angle=-90]{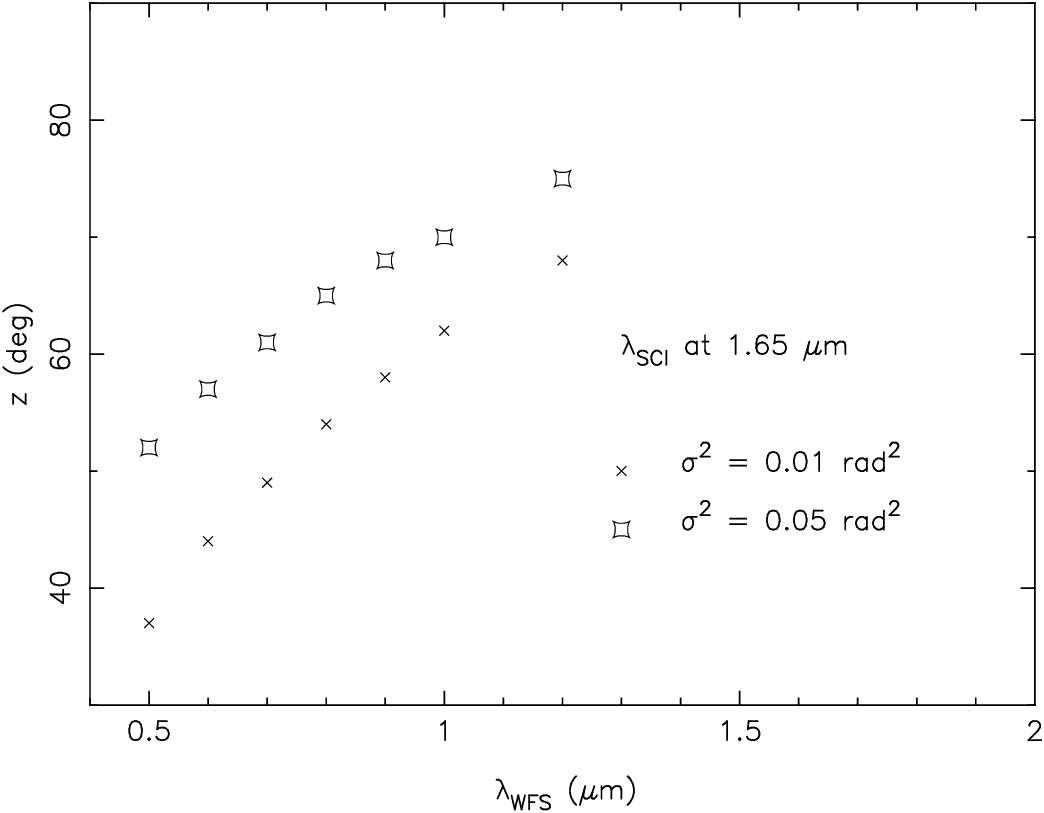}
\caption{Wavelength dependencies of zenith distances giving 
$\sigma^2$ = 0.01 and 0.05.  $\lambda_{SCI} = 1.65 $ $\mu$m.
\label{Z1.65}}
\end{figure}

\clearpage

\begin{figure}[tbp]
\includegraphics[angle=-90]{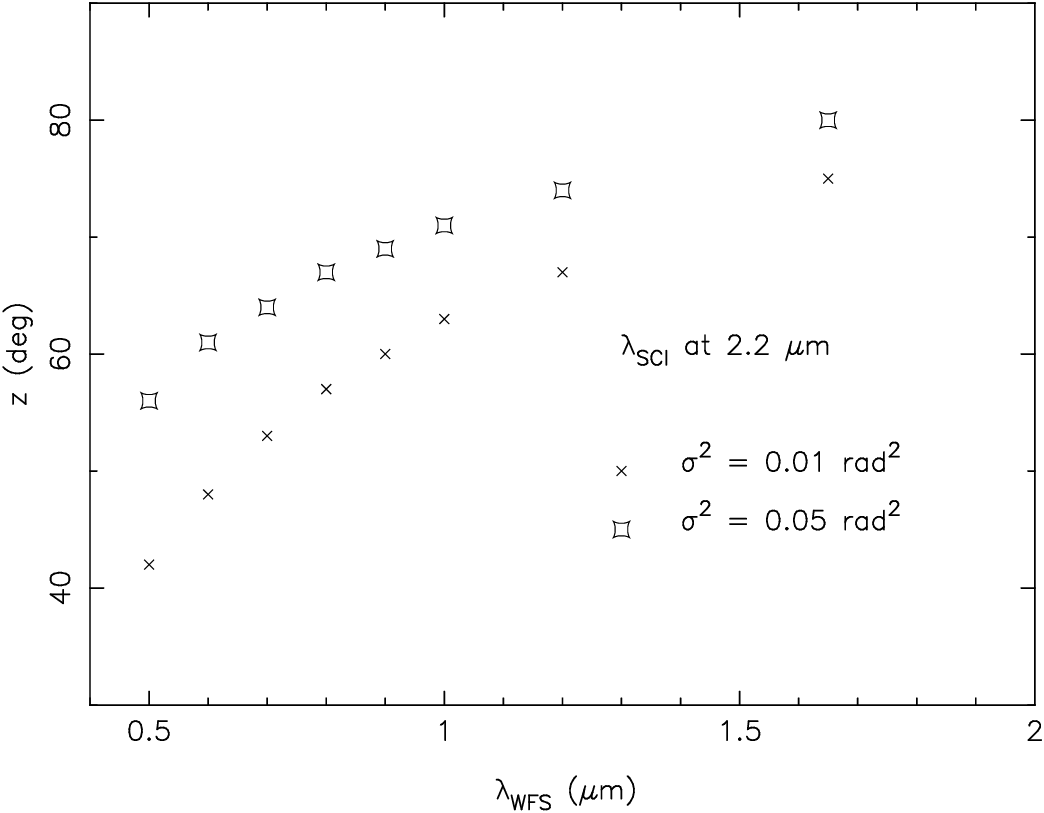}
\caption{Wavelength dependencies of zenith distances giving 
$\sigma^2$ = 0.01 and 0.05.  $\lambda_{SCI} = 2.2 $ $\mu$m.
\label{Z2.2}}
\end{figure}

\end{document}